\documentclass{article}

\usepackage{PRIMEarxiv}

\usepackage[utf8]{inputenc} 
\usepackage[T1]{fontenc}    
\usepackage{hyperref}       
\usepackage{url}            
\usepackage{booktabs}       
\usepackage{amsfonts}       
\usepackage{nicefrac}       
\usepackage{microtype}      
\usepackage{fancyhdr}       
\usepackage{graphicx}       
\graphicspath{{media/}}     
\usepackage{float}

\title{Augmenting Captions with Emotional Cues: An AR Interface for Real-Time Accessible Communication}

\author{
  Sunday David Ubur \\
  Department of Computer Science \\
  Virginia Tech \\
  Blacksburg, VA, USA \\
  \texttt{uburs@vt.edu} \\
}

\begin{document}
\maketitle

\begin{abstract}
This paper introduces an augmented reality (AR) captioning framework designed to support Deaf and Hard of Hearing (DHH) learners in STEM classrooms by integrating non-verbal emotional cues into live transcriptions. Unlike conventional captioning systems that offer only plain text, our system fuses real-time speech recognition with affective and visual signal interpretation—including facial movements, gestures, and vocal tone—to produce emotionally enriched captions. These enhanced captions are rendered in an AR interface developed with Unity, and provide contextual annotations such as speaker tone markers (e.g., "[concerned]") and gesture indicators (e.g., "[nods]"). The system leverages live camera and microphone input, processed through AI models to detect multimodal cues. Findings from preliminary evaluations suggest that this AR-based captioning can significantly enhances comprehension and reduces cognitive effort compared to standard captions. Our work emphasizes the potential of immersive environments for inclusive, emotion-aware educational accessibility.
\end{abstract}

\keywords{
Artificial Intelligence \and 
Augmented Reality \and 
Automatic Speech Recognition \and 
Emotion Recognition \and 
Real-Time Captioning \and 
Accessibility \and 
Multimodal Communication \and 
Deaf and Hard of Hearing \and 
STEM Education \and 
Human-Computer Interaction
}

\section{Introduction}
Ensuring equitable access to STEM education for Deaf and Hard of Hearing (DHH) learners requires more than textual transcription—it demands solutions that reflect the full richness of human communication. While automatic speech recognition (ASR) and Communication Access Realtime Translation (CART) have enabled real-time captioning for classroom content, they often omit vital non-verbal and emotional cues that shape meaning and comprehension \cite{rashid2008representing, ekman1992argument, pang2024cross}. 

Emotional indicators—such as vocal tone, facial expressions, and gestures—help convey speaker intent, stress, urgency, and attitude \cite{ekman1992argument, guyer2021paralinguistic}. In cognitively demanding fields like STEM, these cues can guide attention to complex ideas, signal transitions, or alert students to safety protocols \cite{hwang2016facial, bechtold2023cognitive}. However, mainstream captioning tools remain largely text-centric, presenting challenges for learners who rely on visual-spatial information processing or who have diverse sensory or cognitive preferences \cite{tisdell2017useful, JSLHR-L-18-0185}.

While prior research has explored web-based and video-captioning enhancements that include affective features—such as dynamic text styling, emojis, or embedded prosody \cite{10.1145/3613904.3642258, 10.1145/3544548.3581511}—most implementations are designed for entertainment-based subtitled videos. Limited work has extended these ideas into real-time, classroom-based educational environments. Furthermore, the visual layout of caption displays remains a barrier: learners must frequently shift their attention between speakers and captions, increasing cognitive load \cite{Mayer01012003}. Also, these studies fall short in exploring how such enhancements influence cognitive processing in educational contexts, where understanding and engagement are crucial for effective learning \cite{asee_peer_54067}.

In this paper, we introduce a novel captioning approach that leverages augmented reality (AR) to integrate emotional and multimodal cues into live transcriptions. Built using Unity and deployed through an AR headset interface, our system captures vocal and visual inputs to generate real-time captions enriched with affective annotations (e.g., “[confused tone]”, “[shrug gesture]”). By anchoring captions within the learner’s visual field, our system minimizes attentional shifts and delivers affect-rich content in context.

This work builds upon recent efforts in emotion-enhanced captioning systems \cite{10.1145/3613905.3650987, 10.1145/3613904.3642162}, while expanding the modality and spatial delivery mechanism through immersive AR. We evaluate the cognitive and perceptual benefits of our approach through a preliminary user study and propose new directions for adaptive, real-time captioning in accessible education.

\section{Related Work}

Recent efforts to enhance captioning accessibility have explored multimodal strategies for enriching real-time speech transcriptions with emotional and contextual cues \cite{ ubur2024narrative}. These studies stem from a growing recognition that effective communication—especially in educational settings—requires more than text alone \cite{rashid2008representing, ekman1992argument}.

Early investigations emphasized the role of prosodic and affective signals in improving message clarity and user engagement. For instance, it has been shown that visualizing variations in vocal pitch, rhythm, and intensity can help viewers track speech emphasis and emotion \cite{10.1145/3544548.3581511, 10.1145/3544548.3581130}. These techniques are particularly beneficial for DHH users who rely on visual information to infer speaker intent or urgency.

Work on expressive avatar and captioning systems has also examined the integration of gesture synthesis and visual overlays. Facial and head gestures have been synthesized to mirror affective states in virtual agents \cite{jia2014head}, while sign language translation models have incorporated sign gestures and facial expressions to emphasize emotional clarity \cite{viegas2022including}.

Despite these advances, many captioning solutions remain optimized for asynchronous content—such as videos or social media—rather than live instructional contexts. Emotional captioning in entertainment domains using animated typography and stylistic enhancements has increased emotional salience \cite{rashid2008representing, 10.1145/1279540.1279551}, though such designs may introduce visual overload when applied in real-time. Further work has explored color-coded captions and dynamic prosodic visualizations, with a primary focus on virtual or pre-recorded formats \cite{10.1145/3613904.3642258, 10.1145/3613905.3650987}.

In educational contexts, especially within STEM domains, cognitive load is a central concern. Studies grounded in cognitive load theory have suggested that integrating non-verbal signals into captioning can reduce extraneous processing and improve learner focus \cite{sweller1994cognitive, bechtold2023cognitive}. However, empirical investigations into such integrations under real-time classroom conditions are limited. While group-based captioning systems have been proposed \cite{mcdonnell2024understanding}, the individual cognitive impact of emotional cues in such settings remains underexplored.

Another key limitation in current captioning tools is the lack of personalization. Existing research highlights how fixed-format captions often fail to accommodate neurodiverse attention patterns or sensory preferences. Although keyword highlighting and gesture icons may enhance understanding for some users, others find these features distracting \cite{10.1145/3613904.3642162, 10.1145/3290607.3312921}.

The approach presented in this paper addresses these challenges through an immersive AR design that supports spatially embedded captioning with real-time emotional cues. Unlike prior work focused on flat-screen interfaces, our AR prototype integrates emotion recognition and multimodal annotation into a wearable system, aiming to improve accessibility and comprehension in live, visually dynamic STEM classrooms.

While earlier research has looked into adding nonverbal cues like prosody \cite{10.1145/3544548.3581511}, emotional expressions \cite{10.1145/3613904.3642258}, and body gestures to captioning systems, these initiatives encounter several challenges. Issues such as heightened cognitive load, interpretive complexity, and the potentially overwhelming effect of visual markers like font variations and emojis are significant hurdles. These shortcomings show the necessity for AI-driven methods to interpret nonverbal cues in real-time ASR captions, aiming to enhance cognitive load management, engagement, and understanding, especially for DHH and neurodiverse students, while also enriching their learning experience. One related work that aligns with ours is \cite{9569193}.

\section{Design Motivation and AR System Overview}

\begin{figure}[htbp]
\centering
\includegraphics[width=0.6\textwidth]{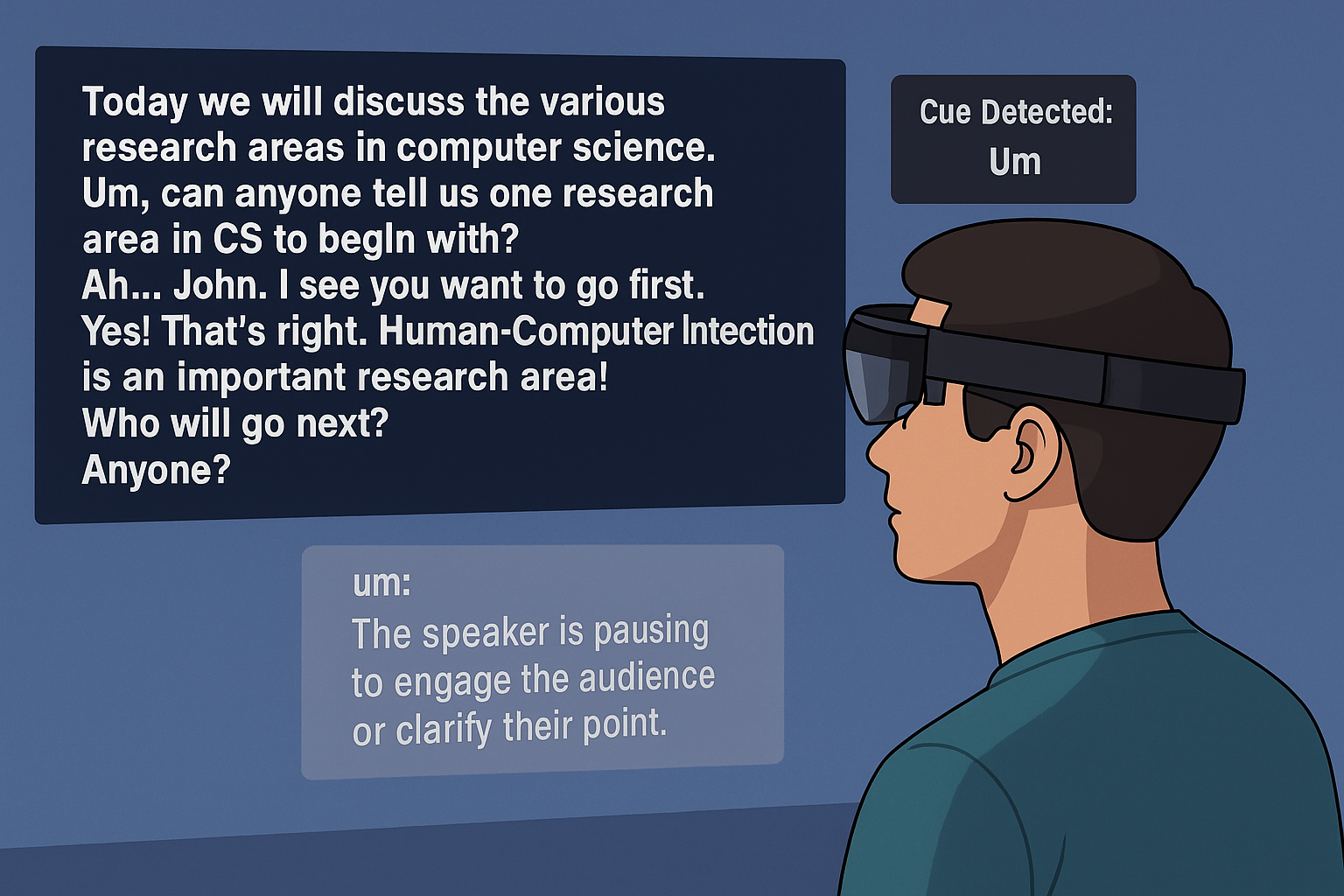}
\caption{Proposed System: An AR real-time captioning system that uses AI to detect multimodal cues and interpret them in real time.}
\label{fig:teaser}
\end{figure}

Our design for an AR captioning system is grounded in prior work involving the development and evaluation of a browser-based prototype that introduced emotion-enhanced captioning for STEM educational content. This earlier tool featured multiple visualization styles, incorporating affective and contextual cues such as tone tags, emojis, and discipline-specific illustrations to supplement real-time transcriptions. Four distinct versions of the captioning interface were tested, each tailored to a specific STEM domain and designed to explore how different multimodal signals—facial expressions, gestures, and vocal intonation—could be represented in caption text.

Insights from this formative study helped shape the transition toward AR. Key challenges identified included limitations of flat-screen interfaces in managing visual attention and the difficulty users faced in simultaneously processing captions, speaker visuals, and instructional content. These findings underscored the need for a more immersive and spatially flexible solution, motivating the creation of an AR-based system capable of integrating multimodal cues directly into a user’s field of view.

The current system advances the earlier prototype by embedding caption streams in a Unity-powered AR application. Live inputs are captured via a camera and microphone, enabling real-time detection of both verbal and non-verbal signals. Speech transcriptions are generated using a lightweight ASR engine, while a machine learning model concurrently analyzes facial movements, gestures, and vocal tone to extract emotional metadata. These data streams are then merged and rendered in an AR interface, where captions can be dynamically annotated with emotion tags (e.g., "[confused]") or gesture indicators (e.g., "[shrugs]").

\begin{figure}[htbp]
\centering
\includegraphics[width=0.6\textwidth]{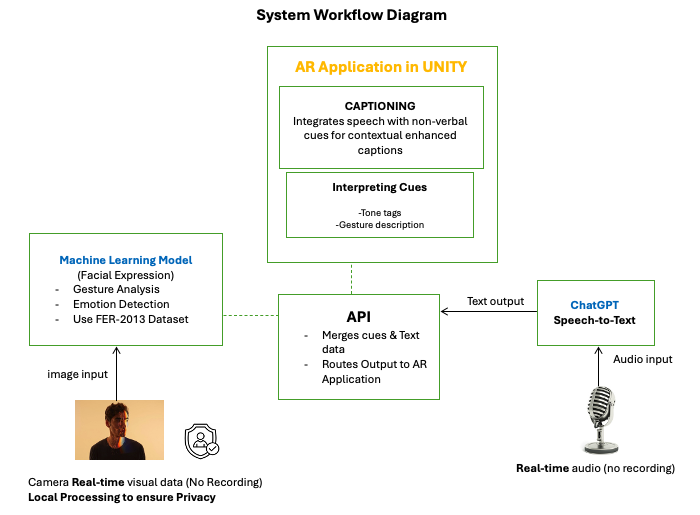} 
\caption{Proposed Prototype Workflow Integration}
\label{fig:system_workflow}
\end{figure}

Developed using Unity 2021 and the Mixed Reality Toolkit 3, the system emphasizes readability, spatial organization, and visual accessibility. Users can adjust features such as caption size, contrast, and placement. Figure~\ref{fig:system_workflow} illustrates the technical architecture of this integrated workflow.

The output is directed to an AR application, built in Unity 2021 using URP and Mixed Reality Toolkit 3, to generate enhanced captions (initial prototype is shown in Figure \ref{fig:teaser}). These captions feature standard speech transcription along with annotations like tone tags (e.g., “[excited]”) or gesture descriptions (e.g., “[pointing gesture]”), and can be customized for visual accessibility. Our emotional enhancements align with findings from XR-based captioning investigations, which emphasize visual clarity and personalization to avoid user fatigue and overload \cite{10.1007/978-3-031-60881-0_24}.

\section{Insights from Prior Evaluation and Motivation for AR Integration}

In a preliminary study conducted prior to this work, we explored how real-time captions could be enhanced through emotional and multimodal cues. A set of four prototype designs were evaluated in an online study with 24 participants, including Deaf, Hard of Hearing, and neurodiverse individuals. Each prototype varied in its inclusion of visual-emotional elements such as facial expression overlays, keyword highlighting, emojis, or discipline-specific visualizations. Quantitative results showed statistically significant reductions in cognitive load—particularly for a minimalist design using text highlighting—while comprehension scores improved compared to a baseline ASR condition. However, designs that incorporated emoji or overlaid captions elicited mixed responses, highlighting the tradeoff between expressiveness and distraction.

These findings, while promising, underscored challenges with flat-screen captioning systems. Participants reported difficulty maintaining attention across multiple screen elements—particularly when captions occluded the speaker’s face or were spatially decoupled from lecture content. Feedback emphasized a strong demand for personalization, with users desiring control over emotive features like emoji and animation. Moreover, caption effectiveness was influenced by user profile (e.g., Deaf vs. ADHD), reinforcing the need for adaptive systems that support diverse accessibility needs.

Building on these insights, our current work transitions from 2D web interfaces to an AR environment, aiming to deliver spatially embedded, emotionally enriched captions that integrate more seamlessly into users’ visual fields. This AR-based approach seeks to reduce the visual fragmentation experienced in prior prototypes by aligning captions with speaker gestures and contextual visuals in real-time, while also offering customizable controls to suit individual preferences.

\section{Discussion}

\paragraph{Reducing Cognitive Load and Enhancing Comprehension}
Findings from the preliminary evaluation of emotion-enhanced captioning interfaces revealed that incorporating affective cues—such as tone tags, visual highlights, and gesture-based annotations—contributed to reduced cognitive load and improved comprehension. These results support foundational principles in Cognitive Load Theory \cite{SWELLER201137}, suggesting that reducing split-attention effects and clarifying speaker intent can enhance learning outcomes. When captions offer cues that help interpret emotional tone or emphasis, learners are less burdened with mentally reconstructing context, which is particularly beneficial in STEM domains where content density is high \cite{bechtold2023cognitive, JSLHR-L-18-0185}.

\paragraph{Balancing Expressiveness with Readability}
Participants expressed a strong appreciation for emotionally expressive captions, particularly those that provided affective context without overwhelming the screen. However, designs that relied on heavy visual embellishments—such as animated emojis or large overlays—were seen as distracting by some users. These findings echo broader concerns in accessibility literature regarding the visual demands of caption presentation, especially for users with ADHD or sensory sensitivities \cite{10.1145/3613904.3642162, rashid2008representing}. This highlights the need for flexible captioning systems that support personalization and adjustable visual density.

\paragraph{Advantages of AR-Based Delivery}
The transition to an AR interface addresses several shortcomings identified in traditional captioning tools \cite{10536263}. Most notably, AR allows captions to be spatially embedded within the user’s visual field, reducing the need for rapid gaze switching between captions and lecture materials. This spatial integration is supported by prior research in multimedia learning, which emphasizes the importance of visual coherence in reducing extraneous cognitive load \cite{Mayer01012003, 10.1007/978-3-031-60881-0_19}. In addition, the AR system supports real-time multimodal processing, aligning emotional annotations (e.g., "[urgent]", "[sarcastic]") with corresponding visual cues such as facial expressions or gestures—an approach that may enhance meaningful construction for both DHH and neurodiverse users \cite{ekman1992argument}.

\paragraph{Design Implications and Accessibility in STEM}
Our results reinforce the need for captioning systems that go beyond literal text transcription to convey emotion, context, and intent. Emotionally enriched captions can help DHH students and others who rely on visual information better interpret speaker affect, follow emphasis cues, and engage with emotionally nuanced academic discourse. These findings are especially relevant in STEM education, where instructors frequently use vocal inflection and gesture to emphasize critical points or signal conceptual transitions \cite{pang2024cross, hwang2016facial}. As multimodal XR tools mature, embedding these cues contextually through AR represents a promising path toward inclusive pedagogy.

\paragraph{Limitations and Future Work}
Although promising, this study is limited in several ways. The evaluation involved a small participant pool and simulated AR outputs, meaning the cognitive and perceptual impacts of full AR deployment remain to be validated in live classroom settings. Furthermore, participants had varying sensory and cognitive profiles, but the system did not yet adapt dynamically to user preferences or neurodiverse needs. 

\begin{itemize}
    \item Future work will involve deploying the emotive-aware system in real-time AR environment, with particular focus on Personalization features,
    \item Cross-cultural interpretations of emotion tags, and
    \item Integration with classroom slide content and instructor gestures.
\end{itemize}

\section{Conclusion}
This paper presents an AR captioning system that enriches real-time speech-to-text output with emotional and multimodal cues. Informed by insights from prior evaluations, the AR interface addresses longstanding challenges in caption readability, cognitive load, and affective interpretation. By spatially anchoring context-aware captions in the user’s field of view and enabling visual customization, this system advances the design of inclusive, emotionally expressive communication tools for education. As captioning systems evolve toward more adaptive and multimodal formats, approaches like  our AR captioning could set the stage for the next generation of accessible learning technologies in STEM and beyond.

\clearpage


\end{document}